\documentclass[epsf,12pt,russian]{article}

\usepackage[dvips]{graphicx}

\begin{document}

\title{Investigation of the anomaly puzzle in
$N=1$ supersymmetric electrodynamics.}

\author{K.V.Stepanyantz \thanks{E-mail:$stepan@phys.msu.ru$}}

\maketitle

\begin{center}
{\em Moscow State University, physical faculty,\\
department of theoretical physics.\\
$119234$, Moscow, Russia}
\end{center}

\begin{abstract}
Using Schwinger-Dyson equations and Ward identities in $N=1$
supersymmetric electrodynamics, regularized by higher derivatives,
we find, that it is possible to calculate some contributions to
the two-point Green function of the gauge field and to the
$\beta$-function exactly to all orders of the perturbation theory.
The results are applied for the investigation of the anomaly
puzzle in the considered theory.
\end{abstract}

%%%%%%%%%%%%%%%%%%%%%%%%%%%%%%%%%%%%%%%%%%%%%%%%%%%%%%%%%%%%%%%%%%%%%%%%%%

\section{Introduction.}
\hspace{\parindent}

The investigation of quantum corrections in supersymmetric
theories is a very interesting and sometimes nontrivial problem.
For example, we remind of the so called "anomaly puzzle". The
matter is that in supersymmetric theories the axial anomaly and
the anomaly of the energy-momentum tensor trace are components of
a single chiral supermultiplet
\cite{Ferrara,Clark,Piquet1,Piquet2}. According to Adler-Bardeen
theorem \cite{Bardeen,Slavnov_Book} the axial anomaly is exhausted
at the one-loop, while the trace anomaly should be proportional to
the $\beta$-function \cite{Adler_Collins} to all orders of the
perturbation theory. Therefore the $\beta$-function in
supersymmetric theories can be supposed to be completely defined
by the one-loop approximation \cite{NSVZ_PL}. However explicit
perturbative calculations in $N=1$ supersymmetric theories
regularized by the dimensional reduction \cite{Siegel} revealed,
that there were higher loops contributions to the $\beta$-function
\cite{Tarasov,Grisaru,Caswell}. Thus we obtain a contradiction,
which was called in the literature "the anomaly puzzle".

Different at the first sight solutions of the anomaly puzzle were
proposed in \cite{SV} and \cite{Arkani}. Nevertheless the results
of both papers were not confirmed by explicit calculations so far
the dimensional reduction was used for the regularization. The
reason, found in \cite{HD_And_DRED}, is the mathematical
inconsistency of the dimensional reduction, which was first
pointed in \cite{Siegel2}. (Note, that unlike the dimensional
reduction, the dimensional regularization \cite{tHV} is not
mathematically inconsistent. However it breaks the supersymmetry
and is, therefore, very inconvenient for using in supersymmetric
theories.) The calculations, made with the mathematically
consistent higher derivative regularization \cite{Slavnov,Bakeyev}
in the two- \cite{hep,tmf2} and three-loop \cite{ThreeLoop}
approximations for the $N=1$ supersymmetric electrodynamics,
showed, that in this case there was no anomaly puzzle. According
to this papers in supersymmetric theories the $\beta$-function,
defined as a derivative of the renormalized coupling constant with
respect to $\ln\mu$, and Gell-Mann-Low function are different due
to the rescaling anomaly. The former is proportional to the trace
anomaly, while the latter has corrections in all orders of the
perturbation theory and in the considered approximation coincides
with the exact Novikov, Shifman, Vainshtein and Zakharov (NSVZ)
$\beta$-function. For the $N=1$ supersymmetric electrodynamics the
exact NSVZ $\beta$-function is:

\begin{equation}\label{NSVZ_Beta}
\beta(\alpha) = \frac{\alpha^2}{\pi}\Big(1-\gamma(\alpha)\Big),
\end{equation}

\noindent where $\gamma(\alpha)$ is the anomalous dimension of the
matter superfield. (First time the exact $\beta$-function was
constructed in \cite{NSVZ_Instanton} as a result of the
investigation of instanton contributions structure.)

Note, that the using of the higher derivative regularization
allows to relate various formal solutions of the anomaly puzzle,
which were proposed in the literature.

Nevertheless, the investigation of the supersymmetric
electrodynamics up to now was restricted by the frames of the
three-loop approximation. However, it would be interesting to
elucidate if it is possible to perform the explicit calculations
exactly to all orders of the perturbation theory. In this paper we
investigate a question, if it is possible to obtain the result,
which is exact to all orders, using Schwinger-Dyson equations and
Ward identities.

The paper is organized as follows:

In Sec. \ref{Section_SUSY_QED} some information about the $N=1$
supersymmetric electrodynamics and its regularization by higher
derivatives is recalled. Schwinger-Dyson equations for the
considered theory are constructed in Sec. \ref{Section_SD}, Ward
identities and their solutions are presented in Sec.
\ref{Section_Ward}. The solutions of Ward identities are
substituted into the Schwinger-Dyson equation for the two-point
Green function of the gauge field in Sec.
\ref{Section_Two_Point_Function}. Then it turned out, that this
Green function in the limit $p\to 0$ can be thus calculated almost
completely. Making a special proposal about the structure of the
remaining contributions it is possible to construct this function
completely (Sec. \ref{Section_Anomaly_Puzzle}). Then divergences
in the two-point Green function are present only in the one-loop
approximation, but Gell-Mann-Low function coincides with the exact
NSVZ $\beta$-function. In the conclusion we discuss the obtained
results.

%%%%%%%%%%%%%%%%%%%%%%%%%%%%%%%%%%%%%%%%%%%%%%%%%%%%%%%%%%%%%%%%%%%

\section{$N=1$ supersymmetric electrodynamics and its
regularization by higher derivatives} \label{Section_SUSY_QED}
\hspace{\parindent}

$N=1$ supersymmetric electrodynamics in the superspace is
described by the following action:

\begin{eqnarray}\label{SQED_Action}
&& S_0 = \frac{1}{4 e^2} \mbox{Re}\int d^4x\,d^2\theta\,W_a C^{ab}
W_b + \frac{1}{4}\int d^4x\, d^4\theta\, \Big(\phi^* e^{2V}\phi
+\tilde\phi^* e^{-2V}\tilde\phi\Big)
+\qquad\nonumber\\
&& + \frac{1}{2}\int d^4x\, d^2\theta\, m\,\tilde\phi\,\phi +
\frac{1}{2}\int d^4x\, d^2\bar\theta\, m \tilde\phi^*\phi^*.
\end{eqnarray}

\noindent Here $\phi$ and $\tilde\phi$ are chiral matter
superfields with the mass $m$, and $V$ is a real scalar
superfield, which contains the gauge field $A_\mu$ as a component.
The superfield $W_a$ is a supersymmetric analog of the stress
tensor of the gauge field. In the Abelian case it is defined by

\begin{equation}
W_a = \frac{1}{16} \bar D (1-\gamma_5) D\Big[(1+\gamma_5)D_a
V\Big],
\end{equation}

\noindent where

\begin{equation}
D = \frac{\partial}{\partial\bar\theta} -
i\gamma^\mu\theta\,\partial_\mu
\end{equation}

\noindent is a supersymmetric covariant derivative.

In order to regularize model (\ref{SQED_Action}) it is possible to
add to its action the term with the higher derivatives:

\begin{eqnarray}\label{Regularized_SQED_Action}
&& S_0 \to S = S_0 + S_{\Lambda} = \frac{1}{4 e^2} \mbox{Re}\int
d^4x\,d^2\theta\,W_a C^{ab} \Big(1+
\frac{\partial^{2n}}{\Lambda^{2n}}\Big) W_b
+\\
&& + \frac{1}{4}\int d^4x\, d^4\theta\, \Big(\phi^* e^{2V}\phi
+\tilde\phi^* e^{-2V}\tilde\phi\Big) + \frac{1}{2}\int d^4x\,
d^2\theta\, m\,\tilde\phi\,\phi + \frac{1}{2}\int d^4x\,
d^2\bar\theta\, m\, \tilde\phi^*\phi^*.\qquad \nonumber
\end{eqnarray}

\noindent It is important to note, that in the Abelian case the
superfield $W^a$ is gauge invariant, so that there are the
ordinary derivatives instead of the covariant ones in the
regularizing term.

The quantization of model (\ref{Regularized_SQED_Action}) can be
made using the standard methods. For this purpose it is convenient
to use the supergraphs technique, described in \cite{West} in
details, and to fix the gauge invariance by adding the following
terms:

\begin{equation}\label{Gauge_Fixing}
S_{gf} = - \frac{1}{64 e^2}\int d^4x\,d^4\theta\, \Bigg(V D^2 \bar
D^2 \Big(1 + \frac{\partial^{2n}}{\Lambda^{2n}}\Big) V + V \bar
D^2 D^2 \Big(1+ \frac{\partial^{2n}}{\Lambda^{2n}}\Big) V\Bigg),
\end{equation}

\noindent where

\begin{equation}
D^2 \equiv \frac{1}{2} \bar D (1+\gamma_5)D;\qquad \bar D^2 \equiv
\frac{1}{2}\bar D (1-\gamma_5) D.
\end{equation}

\noindent After adding such terms a part of the action, quadratic
in the superfield $V$ will have the simplest form

\begin{equation}
S_{gauge} + S_{gf} = \frac{1}{4 e^2}\int d^4x\,d^4\theta\,
V\partial^2 \Big(1+ \frac{\partial^{2n}}{\Lambda^{2n}}\Big) V.
\end{equation}

\noindent In the Abelian case, considered here, diagrams,
containing ghost loops are absent.

It is well known (see e.f. \cite{hep}), that adding of the higher
derivative term does not remove divergences from one-loop
diagrams. In order to regularize them, it is necessary to insert
in the generating functional the Pauli-Villars determinants
\cite{Slavnov_Book}.

Due to the supersymmetric gauge invariance

\begin{equation}\label{Gauge_Transformations}
V \to V - \frac{1}{2}(A+A^+); \qquad \phi\to e^{A}\phi;\qquad
\tilde\phi\to e^{-A} \tilde\phi,
\end{equation}

\noindent where $A$ is an arbitrary chiral scalar superfield, the
renormalized action can be written as

\begin{eqnarray}\label{Renormalized_Action}
&& S_{ren} = \frac{1}{4 e^2} Z_3(e,\Lambda/\mu)\, \mbox{Re}\int
d^4x\,d^2\theta\,W_a C^{ab} \Big(1+
\frac{\partial^{2n}}{\Lambda^{2n}}\Big) W_b
+\nonumber\\
&& + Z(e,\Lambda/\mu)\,\frac{1}{4}\int d^4x\, d^4\theta\,
\Big(\phi^* e^{2V}\phi +\tilde\phi^* e^{-2V}\tilde\phi\Big)
+\nonumber\\
&& \qquad\qquad\qquad\qquad\qquad\qquad + \frac{1}{2}\int d^4x\,
d^2\theta\, m\,\tilde\phi\,\phi + \frac{1}{2}\int d^4x\,
d^2\bar\theta\, m \tilde\phi^*\phi^*, \qquad
\end{eqnarray}

\noindent where we take into account, that due to the
nonrenormalization theorem \cite{West} the mass term is not
renormalized in the perturbation theory. Therefore, the generating
functional can be written as

\begin{equation}\label{Modified_Z}
Z = \int DV\,D\phi\,D\tilde \phi\, \prod\limits_i \Big(\det
PV(V,M_i)\Big)^{c_i}
\exp\Big(i(S_{ren}+S_{gf}+S_S+S_{\phi_0})\Big),
\end{equation}

\noindent where the renormalized action $S_{ren}$ is given by Eq.
(\ref{Renormalized_Action}), the gauge fixing action -- by Eq.
(\ref{Gauge_Fixing}) (It is convenient to substitute $e$ by $e_0$
in Eq. (\ref{Gauge_Fixing}), that we will assume below), the
Pauli-Villars determinants are defined by

\begin{equation}\label{PV_Determinants}
\Big(\det PV(V,M)\Big)^{-1} = \int D\Phi\,D\tilde \Phi\,
\exp\Big(i S_{PV}\Big),
\end{equation}

\noindent where

\begin{eqnarray}
&& S_{PV}\equiv Z(e,\Lambda/\mu) \frac{1}{4} \int
d^4x\,d^4\theta\, \Big(\Phi^* e^{2V}\Phi
+\qquad\nonumber\\
&&\qquad\qquad\qquad + \tilde\Phi^* e^{-2V}\tilde\Phi \Big) +
\frac{1}{2}\int d^4x\,d^2\theta\, M \tilde\Phi \Phi +
\frac{1}{2}\int d^4x\,d^2\bar\theta\, M \tilde\Phi^* \Phi^*,\qquad
\end{eqnarray}

\noindent and the coefficients $c_i$ satisfy conditions

\begin{equation}
\sum\limits_i c_i = 1;\qquad \sum\limits_i c_i M_i^2 = 0.
\end{equation}

\noindent Below we will assume, that $M_i = a_i\Lambda$, where
$a_i$ are some constants. Insertion of the Pauli-Villars
determinants allows to cancel the remaining divergences in all
one-loop diagrams, including diagrams, containing insertions of
counterterms.

The source terms are written as

\begin{eqnarray}\label{Sources}
&& S_S = \int d^4x\,d^4\theta\,J V + \int d^4x\,d^2\theta\,
\Big(j\,\phi + \tilde j\,\tilde\phi \Big) + \int
d^4x\,d^2\bar\theta\, \Big(j^*\phi^* + \tilde j^*
\tilde\phi^*\Big).
\end{eqnarray}

\noindent Moreover, we introduce into generating functional
(\ref{Modified_Z}) the expression

\begin{equation}
S_{\phi_0} = \frac{1}{4}\int d^4x\,d^4\theta\,\Big(\phi_0^*\,
e^{2V} \phi + \phi^*\, e^{2V} \phi_0 + \tilde\phi_0^*\,
e^{-2V}\tilde\phi + \tilde\phi^*\, e^{-2V}\tilde\phi_0 \Big),
\end{equation}

\noindent where $\phi_0$, $\phi_0^*$, $\tilde\phi_0$ and
$\tilde\phi_0^*$ are scalar superfields. They are some parameters,
which are not chiral or antichiral. In principle, it is not
necessary to introduce the term $S_{\phi_0}$ into the generating
functional, but the presence of the parameters $\phi_0$ e t.c.
will be useful for us later for the investigation of
Schwinger-Dyson equations.

In our notations the generating functional for the connected Green
functions is

\begin{equation}\label{W}
W = - i\ln Z,
\end{equation}

\noindent and an effective action is obtained by making a Legendre
transformation:

\begin{equation}\label{Gamma}
\Gamma = W - \int d^4x\,d^4\theta\,J V - \int d^4x\,d^2\theta\,
\Big(j\,\phi + \tilde j\,\tilde\phi \Big) - \int
d^4x\,d^2\bar\theta\, \Big(j^*\phi^* + \tilde j^* \tilde\phi^*
\Big),
\end{equation}

\noindent where $J$, $j$ and $\tilde j$ is to be eliminated in
terms of the fields $V$, $\phi$ and $\tilde\phi$, through solving
equations

\begin{equation}
V = \frac{\delta W}{\delta J};\qquad \phi = \frac{\delta W}{\delta
j};\qquad \tilde\phi = \frac{\delta W}{\delta\tilde j}.
\end{equation}

%%%%%%%%%%%%%%%%%%%%%%%%%%%%%%%%%%%%%%%%%%%%%%%%%%%%%%%%%%%%%%%%%%%

\section{Schwinger-Dyson equations for $N=1$ supersymmetric
electrodynamics} \label{Section_SD} \hspace{\parindent}

We will try to calculate the two-point Green function of the gauge
field. First for the simplicity we set the renormalization
constant $Z$ in Eq. (\ref{Renormalized_Action}) equal to 1, that
corresponds to taking into account all diagrams, which do not
contain insertions of counterterms on lines of the matter
superfields. Later we will see, that if $Z\ne 1$, the two-point
Green function of the gauge field can be easily found from the
corresponding result with $Z=1$.

First we should write Schwinger-Dyson equations for the considered
theory ($Z=1$). For this purpose first we split the action into
two parts:

\begin{equation}
S = S_2 + S_I,
\end{equation}

\noindent where

\begin{eqnarray}
&& S_2 \equiv \frac{1}{4 e_0^2}\int d^4x\,d^4\theta\, V\partial^2
\Big(1+ \frac{\partial^{2n}}{\Lambda^{2n}}\Big) V +
\frac{1}{4}\int d^4x\,d^4\theta\,\Big(\phi^* \phi + \tilde\phi^*
\tilde\phi\Big) +\nonumber\\
&&\qquad\qquad\qquad\qquad\qquad\qquad\qquad + \frac{1}{2}\int
d^4x\, d^2\theta\, m\,\tilde\phi\,\phi + \frac{1}{2}\int d^4x\,
d^2\bar\theta\, m \tilde\phi^*\phi^*\qquad
\end{eqnarray}

\noindent is a quadratic part of the action (including gauge
fixing terms) and

\begin{equation}\label{S_I}
S_I \equiv \frac{1}{4}\int d^4x\,d^4\theta\,\Big(\phi^* (e^{2V}-1)
\phi + \tilde\phi^* (e^{-2V}-1)\tilde\phi\Big)
\end{equation}

\noindent is the interaction.

Then, setting all fields $\phi_0 = 0$, generating functional
(\ref{Modified_Z}) can be written as

\begin{eqnarray}
&& Z[J,j]\Bigg|_{\phi_0 = 0} = \prod\limits_i \Big(\det
PV\Big(\frac{1}{i}\frac{\delta}{\delta J},M_i\Big)\Big)^{c_i}
\exp\Bigg(iS_I\Big[\frac{1}{i}\frac{\delta}{\delta
J},\frac{1}{i}\frac{\delta}{\delta j}\Big]\Bigg)\times\nonumber\\
&&\times \int DV D\phi \exp\Bigg(i\,S_2(V,\phi)+i\,
S_S\Bigg)\Bigg|_{\phi_0 = 0},
\end{eqnarray}

\noindent where from all matter superfields we wrote explicitly
only the field $\phi$ for the brevity of the notations. (Below we
will also omit a condition $\phi_0 = 0$ for the same reasons.)

The remaining integral in this expression is gaussian and can be
easily calculated:

\begin{equation}\label{Perturbative_Z}
Z[J,j] = \prod\limits_i \Big(\det
PV\Big(\frac{1}{i}\frac{\delta}{\delta J},M_i\Big)\Big)^{c_i}
\exp\Bigg(iS_I\Big[\frac{1}{i}\frac{\delta}{\delta
J},\frac{1}{i}\frac{\delta}{\delta j}\Big]\Bigg) Z_0,
\end{equation}

\noindent where

\begin{eqnarray}
&& Z_0 = \exp\Bigg\{i \int d^4x\,d^4\theta\Bigg(- J
\frac{e_0^2}{\partial^2 \Big(1+\partial^{2n}/\Lambda^{2n}\Big)} J
+ j^* \frac{1}{\partial^2+m^2}  j + \tilde j^*
\frac{1}{\partial^2+m^2} \tilde j +\nonumber\\
&& + j \frac{m}{\partial^2+m^2}\, \frac{D^2}{4\partial^2}\, \tilde
j + j^* \frac{m}{\partial^2+m^2}\, \frac{\bar D^2}{4\partial^2}\,
\tilde j^* \Bigg)\Bigg\}.
\end{eqnarray}

Let us differentiate this expression with respect to $J_x$ (an
index $x$ here and later denotes the argument of a function):

\begin{eqnarray}
&& \frac{\delta Z}{\delta J_x} = - \prod\limits_i \Big(\det
PV\Big(\frac{1}{i}\frac{\delta}{\delta J},M_i\Big)\Big)^{c_i}
\exp\Bigg(i S_I\Big[\frac{1}{i}\frac{\delta}{\delta
J},\frac{1}{i}\frac{\delta}{\delta j}\Big]\Bigg)
\frac{2ie_0^2}{\partial^2 \Big(1+\partial^{2n}/\Lambda^{2n}\Big)}
J_x
Z_0 =\nonumber\\
&& = -\frac{2ie_0^2}{\partial^2
\Big(1+\partial^{2n}/\Lambda^{2n}\Big)} \Bigg(J_x + \sum\limits_i
c_i \frac{\delta}{\delta V_x}\ln \det
PV\Big(\frac{1}{i}\frac{\delta}{\delta J},M_i\Big) + \frac{\delta
S_I}{\delta V_x}\Bigg[\frac{1}{i}\frac{\delta}{\delta
J},\frac{1}{i}\frac{\delta}{\delta j}\Bigg]\Bigg) Z,\nonumber\\
\end{eqnarray}

\noindent where

\begin{equation}\label{SI}
\frac{\delta S_I}{\delta V_x}\Big[V,\phi\Big] =
\frac{1}{2}\Big(\phi^*_x e^{2V_x} \phi_x - \tilde\phi^*_x
e^{-2V_x} \tilde\phi_x\Big).
\end{equation}

\noindent according to Eq. (\ref{S_I}).

Taking into account that $Z=\exp(iW)$, we rewrite this identity in
terms of the functional $W$ as follows:

\begin{eqnarray}
&& \frac{\delta W}{\delta J_x} = -\frac{2e_0^2}{\partial^2
\Big(1+\partial^{2n}/\Lambda^{2n}\Big)} \Bigg(J_x + \sum\limits_i
c_i \frac{\delta}{\delta V_x}\ln \det
PV\Big(\frac{1}{i}\frac{\delta}{\delta J}+\frac{\delta W}{\delta
J},M_i\Big) +\nonumber\\
&& + \frac{\delta S_I}{\delta
V_x}\Bigg[\frac{1}{i}\frac{\delta}{\delta J}  +\frac{\delta
W}{\delta J},\frac{1}{i}\frac{\delta}{\delta j}+\frac{\delta
W}{\delta j}\Bigg]\Bigg).
\end{eqnarray}

Passing at last to the effective action $\Gamma$, we obtain the
following identity:

\begin{eqnarray}\label{SD1}
&& \frac{\delta\Gamma}{\delta V_x} = \frac{1}{2e_0^2}\partial^2
\Big(1+\partial^{2n}/\Lambda^{2n}\Big) V_x + \sum\limits_i c_i
\frac{\delta}{\delta V_x}\ln \det
PV\Big(\frac{1}{i}\frac{\delta}{\delta J}+V,M_i\Big)
+\nonumber\\
&& + \frac{\delta S_I}{\delta
V_x}\Bigg[\frac{1}{i}\frac{\delta}{\delta J}+V,
\frac{1}{i}\frac{\delta}{\delta j}+\phi\Bigg].
\end{eqnarray}

\noindent Here the derivatives with respect to the sources should
be expressed in terms of derivatives with respect to the fields.
For example, taking into account that

\begin{equation}
\frac{\delta \phi_x}{\delta\phi_y} = - \frac{1}{2} D^2
\delta^8_{xy};\qquad \frac{\delta \phi^*_x}{\delta\phi^*_y} = -
\frac{1}{2} D^2 \delta^8_{xy}
\end{equation}

\noindent where

\begin{equation}
\delta^8_{xy} \equiv \delta^4(x-y) \delta^4(\theta_x-\theta_y),
\end{equation}

\noindent we obtain

\begin{eqnarray}
&& \frac{\delta}{\delta j^*_x} = \int d^8z\,\Bigg(\frac{\delta
\phi_z}{\delta j^*_x}\,\frac{D^2}{8\partial^2}\frac{\delta}{\delta
\phi_z} + \frac{\delta \phi_z^*}{\delta
j^*_x}\,\frac{D^2}{8\partial^2}\frac{\delta}{\delta \phi_z^*} +
\frac{\delta \tilde \phi_z}{\delta
j^*_x}\,\frac{D^2}{8\partial^2}\frac{\delta}{\delta \tilde \phi_z}
+ \frac{\delta \tilde \phi_z^*}{\delta
j^*_x}\,\frac{D^2}{8\partial^2}\frac{\delta}{\delta
\tilde\phi_z^*}
+\nonumber\\
&& + \frac{\delta V_z}{\delta j^*_x}\frac{\delta}{\delta
V_z}\Bigg) = \int d^8z\,\Bigg[\Bigg(\frac{\delta^2\Gamma}{\delta
\phi_z \delta \phi^*_x}\Bigg)^{-1}
\,\frac{D^2}{8\partial^2}\frac{\delta}{\delta \phi_z} +
\Bigg(\frac{\delta^2\Gamma}{\delta \phi_z^* \delta
\phi^*_x}\Bigg)^{-1}\,\frac{D^2}{8\partial^2}\frac{\delta}{\delta
\phi_z^*} +\nonumber\\
&& + \Bigg(\frac{\delta^2\Gamma}{\delta \tilde \phi_z \delta
\phi^*_x}\Bigg)^{-1} \,\frac{D^2}{8\partial^2}\frac{\delta}{\delta
\tilde \phi_z} + \Bigg(\frac{\delta^2\Gamma}{\delta \tilde
\phi_z^* \delta \phi^*_x}\Bigg)^{-1}
\,\frac{D^2}{8\partial^2}\frac{\delta}{\delta \tilde\phi_z^*}+
\Bigg(\frac{\delta^2\Gamma}{\delta V_z \delta
\phi^*_x}\Bigg)^{-1}\frac{\delta}{\delta V_z}\Bigg],
\end{eqnarray}

\noindent where

\begin{equation}
\int d^8z \equiv \int d^4z\,d^4\theta_z.
\end{equation}

\noindent It is important to note, that if all fields are set
equal to 0, nontrivial contributions come only from the first and
the fourth terms due to the continuousness of matter superfields
lines (if $m=0$ only the first term gives a nontrivial
contribution).

Let us substitute the explicit expression for $S_I$, given by Eq.
(\ref{SI}), into Eq. (\ref{SD1}), differentiate the result with
respect to $V_y$ and then set all fields equal to 0:

\begin{eqnarray}\label{Delta_Gamma_VV}
&& \frac{\delta^2\Gamma}{\delta V_x\,\delta V_y}\Bigg|_{V,\phi=0}
= \frac{1}{2e_0^2}\partial^2
\Big(1+\partial^{2n}/\Lambda^{2n}\Big) \delta^8_{xy} +
\sum\limits_i c_i \frac{\delta^2}{\delta V_x \delta V_y}\ln \det
PV\Big(\frac{1}{i}\frac{\delta}{\delta
J}+V,M_i\Big) +\nonumber\\
&& + \frac{\delta}{\delta V_y} \int
d^8x_1\,\Bigg(\frac{\delta^2\Gamma}{\delta\phi^*_x\delta
\phi_1}\Bigg)^{-1}\frac{D_1^2}{16i\partial^2}\frac{\delta}{\delta
\phi_1} \exp\Bigg(\frac{2}{i}\frac{\delta}{\delta J_x}+2V_x\Bigg)
\phi_x +\nonumber\\
&& + \frac{\delta}{\delta V_y} \int
d^8x_1\,\Bigg(\frac{\delta^2\Gamma}{\delta\phi^*_x\delta \tilde
\phi^*_1}\Bigg)^{-1}\frac{\bar
D_1^2}{16i\partial^2}\frac{\delta}{\delta \tilde\phi^*_1}
\exp\Bigg(\frac{2}{i}\frac{\delta}{\delta J_x}+2V_x\Bigg)
\phi_x -\nonumber\\
&& - \frac{\delta}{\delta V_y} \int
d^8x_1\,\Bigg(\frac{\delta^2\Gamma}{\delta\tilde\phi^*_x\delta
\tilde\phi_1}\Bigg)^{-1}\frac{D^2}{16i\partial^2}
\frac{\delta}{\delta \tilde\phi_1}
\exp\Bigg(-\frac{2}{i}\frac{\delta}{\delta J_x}-2V_x\Bigg)
\tilde\phi_x - \nonumber\\
&& - \frac{\delta}{\delta V_y} \int
d^8x_1\,\Bigg(\frac{\delta^2\Gamma}{\delta\tilde\phi^*_x\delta
\phi_1^*}\Bigg)^{-1}\frac{\bar D^2}{16i\partial^2}
\frac{\delta}{\delta \phi_1^*}
\exp\Bigg(-\frac{2}{i}\frac{\delta}{\delta J_x}-2V_x\Bigg)
\tilde\phi_x\Bigg|_{V,\phi=0}.
\end{eqnarray}

\noindent Using the Leibnitz rule this expression can be presented
in the following form:

\begin{eqnarray}\label{Main_Identity}
&& \frac{\delta^2\Gamma}{\delta V_x\,\delta V_y}\Bigg|_{V,\phi=0}
= \frac{1}{2e_0^2}\partial^2
\Big(1+\partial^{2n}/\Lambda^{2n}\Big) \delta^8_{xy} +
\sum\limits_i c_i \frac{\delta^2}{\delta V_x \delta V_y}\ln \det
PV\Big(\frac{1}{i}\frac{\delta}{\delta
J}+V,M_i\Big) +\nonumber\\
&& + \int d^8x_1\,\Bigg(\frac{\delta^2\Gamma}{\delta\phi^*_x\delta
\phi_1}\Bigg)^{-1} \frac{D_1^2}{16i\partial^2}
\frac{\delta}{\delta V_y} \frac{\delta}{\delta
\phi_1}\exp\Bigg(\frac{2}{i}\frac{\delta}{\delta J_x}+2V_x\Bigg)
\phi_x - \int d^8x_1\,d^8x_2\,d^8x_3\,
\times\nonumber\\
&&\times \Bigg\{ \Bigg(\frac{\delta^2\Gamma}{\delta\phi^*_x\delta
\phi_2}\Bigg)^{-1} \frac{D_2^2}{8\partial_2^2} \frac{\bar
D_3^2}{8\partial_3^2} \frac{\delta^3\Gamma}{\delta V_y
\delta\phi_2\delta \phi^*_3}
\Bigg(\frac{\delta^2\Gamma}{\delta\phi^*_3\delta
\phi_1}\Bigg)^{-1} +
\Bigg(\frac{\delta^2\Gamma}{\delta\phi^*_x\delta \tilde
\phi_2^*}\Bigg)^{-1} \frac{\bar D_2^2}{8\partial_2^2}
\frac{D_3^2}{8\partial_3^2}
\times\nonumber\\
&& \times \frac{\delta^3\Gamma}{\delta V_y \delta\tilde\phi_2^*
\delta \tilde\phi_3}
\Bigg(\frac{\delta^2\Gamma}{\delta\tilde\phi_3\delta
\phi_1}\Bigg)^{-1} +
\Bigg(\frac{\delta^2\Gamma}{\delta\phi^*_x\delta
\phi_2}\Bigg)^{-1} \frac{D_2^2}{8\partial_2^2}
\frac{D_3^2}{8\partial_3^2} \frac{\delta^3\Gamma}{\delta V_y
\delta\phi_2\delta \tilde\phi_3}
\Bigg(\frac{\delta^2\Gamma}{\delta\tilde\phi_3\delta
\phi_1}\Bigg)^{-1}
+\nonumber\\
&& + \Bigg(\frac{\delta^2\Gamma}{\delta\phi^*_x\delta\tilde
\phi_2^*}\Bigg)^{-1} \frac{\bar D_2^2}{8\partial_2^2} \frac{\bar
D_3^2}{8\partial_3^2} \frac{\delta^3\Gamma}{\delta V_y
\delta\tilde\phi_2^*\delta \phi^*_3}
\Bigg(\frac{\delta^2\Gamma}{\delta\phi^*_3\delta
\phi_1}\Bigg)^{-1}\Bigg\} \frac{D_1^2}{16i\partial_1^2}
\frac{\delta}{\delta \phi_1}
\exp\Bigg(\frac{2}{i}\frac{\delta}{\delta J_x}+2V_x\Bigg) \phi_x
+\nonumber\\
&& + \,\mbox{the other similar terms} \Bigg|_{V,\phi=0}
\end{eqnarray}

This equation is a Schwinger-Dyson equation for the considered
theory. It is more convenient to present it in the graphical form,
as a sum of two diagrams, presented in Fig.
\ref{Figure_Effective_Diagrams}. The terms, which contain the
integration over $d^8x_1$ only, correspond to the second diagram
in this figure, while the terms, containing integration over
$d^8x_1\,d^8x_2\,d^8x_3$, -- to the first diagram. Here the double
lines denote the exact propagator, the large circle -- the
effective vertex and two adjacent circles -- the effective vertex,
consisting of 1PI diagrams, in which one of the external lines is
attached to the very left edge.

Contributions of the Pauli-Villars determinants are calculated
similarly. It is convenient to introduce the notation

\begin{equation}\label{Z_PV}
Z_{PV}[V,j_{PV},M_i] \equiv \int D\Phi\,D\tilde \Phi\, \exp\Bigg(i
S_{PV} + i S_{S_{PV}} + i S_{\Phi_0}\Big),
\end{equation}

\noindent where

\begin{eqnarray}\label{PV_Sources}
&& S_{S_{PV}} = \int d^4x\,d^2\theta\, \Big(j_{PV}\,\Phi + \tilde
j_{PV}\,\tilde\Phi \Big) + \int d^4x\,d^2\bar\theta\,
\Big(j_{PV}^*\Phi^* + \tilde j_{PV}^*
\tilde\Phi^*\Big);\nonumber\\
&& S_{\Phi_0} = \frac{1}{4}\int d^4x\,d^4\theta\,\Big(\Phi_0^*\,
e^{2V} \Phi + \Phi^*\, e^{2V} \Phi_0 + \tilde\Phi_0^*\,
e^{-2V}\tilde\Phi + \tilde\Phi^*\, e^{-2V}\tilde\Phi_0\Big).
\end{eqnarray}

\noindent Then according to Eq. (\ref{PV_Determinants}) the
determinants, which is necessary for us, can be written as

\begin{equation}
\Big(\det PV(V,M_i)\Big)^{-1} =
Z_{PV}[V,j_{PV},M_i]\Bigg|_{\Phi_0,j_{PV}=0}.
\end{equation}

\noindent Moreover, let us define functionals

\begin{eqnarray}
&& W_{PV}[V,j_{PV},M_i] \equiv -i \ln
Z_{PV}[V,j_{PV},M_i];\nonumber\\
&& \Gamma_{PV}[V,\Phi,M_i] = W_{PV} - \int d^4x\,d^2\theta\,
\Big(j_{PV}\,\Phi + \tilde j_{PV}\,\tilde\Phi \Big)
-\nonumber\\
&&\qquad\qquad\qquad\qquad\qquad\qquad\qquad\qquad - \int
d^4x\,d^2\bar\theta\, \Big(j_{PV}^*\Phi^* + \tilde j_{PV}^*
\tilde\Phi^* \Big),\qquad
\end{eqnarray}

\noindent where the fields $\Phi$ and $\tilde\Phi$ are related
with the sources $j_{PV}$ and $\tilde j_{PV}$ by equations

\begin{equation}
\Phi = \frac{\delta W_{PV}}{\delta j_{PV}};\qquad \tilde\Phi =
\frac{\delta W_{PV}}{\delta\tilde j_{PV}}.
\end{equation}

\noindent Differentiating Eq. (\ref{Z_PV}) with respect to $V_x$
we find

\begin{eqnarray}
&& \frac{\delta Z_{PV}}{\delta V_x} = - \frac{i}{2}
\Bigg(\frac{\delta}{\delta j^*_{x\,PV}} e^{2V_x}
\frac{\delta}{\delta j_{x\,PV}} - \frac{\delta}{\delta \tilde
j^*_{x\,PV}} e^{-2V_x} \frac{\delta}{\delta \tilde j_{x\,PV}}
\Bigg) Z_{PV}. \qquad
\end{eqnarray}

\noindent Dividing this expression by $Z_{PV}$ and setting $\Phi$
and $\tilde\Phi$ equal to 0, we obtain

\begin{equation}
\frac{\delta W_{PV}}{\delta V_x}\Bigg|_{\Phi=0} = - \frac{1}{2}
\frac{\delta}{\delta j^*_{x\,PV}} e^{2V_x} \frac{\delta
W_{PV}}{\delta j_{x\,PV}} + \frac{1}{2} \frac{\delta}{\delta
\tilde j^*_{x\,PV}} e^{-2V_x} \frac{\delta W_{PV}}{\delta \tilde
j_{x\,PV}}\Bigg|_{\Phi=0}.
\end{equation}

\noindent In terms of $\Gamma_{PV}$ this identity can be written
as

\begin{equation}
\frac{\delta \Gamma_{PV}}{\delta V_x}\Bigg|_{\Phi=0} = -
\frac{1}{2} \frac{\delta}{\delta j^*_{x\,PV}} e^{2V_x} \Phi_x +
\frac{1}{2} \frac{\delta}{\delta \tilde j^*_{x\,PV}} e^{-2V_x}
\tilde\Phi_x\Bigg|_{\Phi=0}.
\end{equation}

\noindent Passing in this equation from the derivatives with
respect to the sources to the derivatives with respect to the
fields we find, that the expression for

\begin{eqnarray}
\frac{\delta^2}{\delta V_x \delta V_y}\ln \det
PV\Big(\frac{1}{i}\frac{\delta}{\delta
J}+V,M_i\Big)\Bigg|_{V,\Phi=0}
\end{eqnarray}

\noindent coincides with the last four terms in Eq.
(\ref{Delta_Gamma_VV}) up to the substitutions $\phi\to \Phi$,
$\tilde\phi\to\tilde\Phi$, $\phi^*\to\Phi^*$ and
$\tilde\phi^*\to\tilde\Phi^*$. Thus we conclude, that the
structure of terms, coming from the Pauli-Villars determinants, is
actually the same as the structure of the other terms and their
calculation is made similarly.

We also need Schwinger-Dyson equations of another type. In order
to derive them let us differentiate the generating functional with
respect to $\phi_0^*$. Then

\begin{equation}
\frac{\delta Z}{\delta\phi_{0z}^*} =
\frac{1}{4}\exp\Bigg(\frac{2}{i}\frac{\delta}{\delta J_z}\Bigg)
\frac{\delta Z}{\delta j_z}.
\end{equation}

\noindent Dividing this equation by $Z$, we find

\begin{equation}
\frac{\delta W}{\delta\phi_{0z}^*} =
\frac{1}{4}\exp\Bigg(\frac{2}{i}\frac{\delta}{\delta J_z} +
2\frac{\delta W}{\delta J_z}\Bigg) \frac{\delta W}{\delta j_z}.
\end{equation}

\noindent Because the field $\phi_0^*$ is a parameter of the
effective action, we obtain the identity

\begin{equation}
\frac{\delta\Gamma}{\delta \phi_{0z}^*} = \frac{\delta
W}{\delta\phi_{0z}^*} =
\frac{1}{4}\exp\Bigg(\frac{2}{i}\frac{\delta}{\delta J_z} + 2
V_z\Bigg) \phi_z.
\end{equation}

\noindent Differentiating this equation and setting then all
fields equal to 0, we find the following equalities, which will be
useful for us later:

\begin{eqnarray}\label{Useful_Identities}
&& \frac{\delta^2\Gamma}{\delta\phi_y \delta \phi_{0z}^*} =
\frac{1}{4} \frac{\delta}{\delta \phi_y}
\exp\Bigg(\frac{2}{i}\frac{\delta}{\delta J_z} + 2 V_z\Bigg)
\phi_z;\nonumber\\
&& \frac{\delta^2\Gamma}{\delta\tilde\phi_y^* \delta \phi_{0z}^*}
= \frac{1}{4} \frac{\delta}{\delta \tilde\phi_y^*}
\exp\Bigg(\frac{2}{i}\frac{\delta}{\delta J_z} + 2
V_z\Bigg) \phi_z;\nonumber\\
&& \frac{\delta^3\Gamma}{\delta V_x \delta\phi_y \delta
\phi_{0z}^*} = \frac{1}{4} \frac{\delta}{\delta V_x}
\frac{\delta}{\delta \phi_y}
\exp\Bigg(\frac{2}{i}\frac{\delta}{\delta J_z} + 2 V_z\Bigg)
\phi_z;\nonumber\\
&& \frac{\delta^3\Gamma}{\delta V_x \delta\tilde\phi_y^* \delta
\phi_{0z}^*} = \frac{1}{4} \frac{\delta}{\delta V_x}
\frac{\delta}{\delta \tilde\phi_y^*}
\exp\Bigg(\frac{2}{i}\frac{\delta}{\delta J_z} + 2 V_z\Bigg)
\phi_z.
\end{eqnarray}

Let us relate these expressions with the ordinary Green functions.
For this purpose it is possible to use other Schwinger-Dyson
equations, which can be derived similarly to those, which were
described earlier (differentiating $Z$ with respect to $j^*$ and
$\tilde j$):

\begin{eqnarray}\label{SD_For_Matter1}
&& (\partial^2+m^2)\phi_x^* = \frac{D_x^2}{2}
\Bigg(\frac{\delta\Gamma}{\delta \phi_x} + \frac{\bar
D_x^2}{8}\Bigg(\Bigg[\exp\Bigg(\frac{2}{i}\frac{\delta}{\delta
J_x}+2V_x\Bigg)-1\Bigg] \phi_x^*\Bigg)
\Bigg) +\nonumber\\
&&\qquad\qquad\qquad\qquad\qquad + 2m
\Bigg(\frac{\delta\Gamma}{\delta \tilde\phi_x^*} +
\frac{D_x^2}{8}\Bigg(
\Bigg[\exp\Bigg(-\frac{2}{i}\frac{\delta}{\delta J_x}-2V_x\Bigg)
-1\Bigg]\tilde\phi_x \Bigg)\Bigg);\qquad
\end{eqnarray}

\begin{eqnarray}\label{SD_For_Matter2}
&& (\partial^2+m^2)\tilde\phi_x = \frac{\bar D_x^2}{2}
\Bigg(\frac{\delta\Gamma}{\delta \tilde\phi_x^*} +
\frac{D_x^2}{8}\Bigg(\Bigg[\exp\Bigg(-\frac{2}{i}\frac{\delta}{\delta
J_x}-2V_x\Bigg)-1\Bigg] \tilde\phi_x\Bigg)
\Bigg) +\nonumber\\
&&\qquad\qquad\qquad\qquad\qquad + 2m
\Bigg(\frac{\delta\Gamma}{\delta \phi_x} + \frac{\bar
D_x^2}{8}\Bigg( \Bigg[\exp\Bigg(\frac{2}{i}\frac{\delta}{\delta
J_x}+2V_x\Bigg) -1\Bigg]\phi_x^* \Bigg)\Bigg).\qquad
\end{eqnarray}

\noindent Differentiating Eqs. (\ref{SD_For_Matter1}) and
(\ref{SD_For_Matter2}) with respect to $\phi_y^*$, and then
setting all fields equal to 0, we obtain a system of equations

\begin{eqnarray}\label{Equation_For_Exponent1}
&& - m^2 D_x^2 \delta^8_{xy} = D_x^2
\Bigg(\frac{\delta^2\Gamma}{\delta\phi_y^* \delta \phi_x} +
\frac{\delta}{\delta\phi_y^*} \frac{\bar D_x^2}{8}
\exp\Bigg(\frac{2}{i}\frac{\delta}{\delta J_x}+2V_x\Bigg) \phi_x^*
\Bigg) +\nonumber\\
&&\qquad\qquad\qquad\qquad\qquad + 4m
\Bigg(\frac{\delta^2\Gamma}{\delta\phi_y^* \delta \tilde\phi_x^*}
+ \frac{D_x^2}{8} \frac{\delta}{\delta\phi_y^*}
\exp\Bigg(-\frac{2}{i}\frac{\delta}{\delta J_x}-2V_x\Bigg)
\tilde\phi_x \Bigg);\qquad
\end{eqnarray}

\begin{eqnarray}\label{Equation_For_Exponent2}
&& 0 = \bar D_x^2 \Bigg(\frac{\delta^2\Gamma}{\delta\phi_y^*
\delta \tilde\phi_x^*} + \frac{\delta}{\delta\phi_y^*}
\frac{D_x^2}{8} \exp\Bigg(-\frac{2}{i}\frac{\delta}{\delta
J_x}-2V_x\Bigg) \tilde\phi_x
\Bigg) +\nonumber\\
&&\qquad\qquad\qquad\qquad\qquad + 4m
\Bigg(\frac{\delta^2\Gamma}{\delta\phi_y^* \delta \phi_x} +
\frac{\bar D_x^2}{8} \frac{\delta}{\delta\phi_y^*}
\exp\Bigg(\frac{2}{i}\frac{\delta}{\delta J_x}+2V_x\Bigg) \phi_x^*
\Bigg).\qquad
\end{eqnarray}

\noindent After simple transformations from Eqs.
(\ref{Equation_For_Exponent1}) and (\ref{Equation_For_Exponent2})
we find

\begin{eqnarray}\label{Exponents}
&& \frac{\delta}{\delta \phi_1}
\exp\Bigg(\frac{2}{i}\frac{\delta}{\delta J_x}+2V_x\Bigg) \phi_x =
\frac{\bar D_x^2}{2\partial^2}
\frac{\delta^2\Gamma}{\delta\phi_1 \delta\phi_x^*};\nonumber\\
&& \frac{\delta}{\delta \tilde \phi_1^*}
\exp\Bigg(\frac{2}{i}\frac{\delta}{\delta J_x}+2V_x\Bigg) \phi_x =
\frac{\bar D_x^2}{2\partial^2}
\Bigg(\frac{\delta^2\Gamma}{\delta\tilde \phi_1^* \delta \phi_x^*}
+ \frac{m}{4} D_x^2\delta^8_{1x} \Bigg).
\end{eqnarray}

Taking into account that a variational derivative with respect to
a chiral superfield is a chiral superfield again, and using
dimensional arguments, we find, that the explicit form of the
two-point Green functions for the matter superfields is

\begin{equation}\label{Explicit_Green_Functions}
\frac{\delta^2\Gamma}{\delta\phi_x^*\delta\phi_y} = \frac{D_x^2
\bar D_x^2}{16} G(\partial^2) \delta^8_{xy};\qquad
\frac{\delta^2\Gamma}{\delta\phi_x\delta\tilde\phi_y} = -
\frac{\bar D_x^2}{4} m J(\partial^2) \delta^8_{xy}.
\end{equation}

\noindent Note, that the corresponding inverse functions, which
are determined from equations

\begin{eqnarray}\label{Equations_For_Inverse_Functions}
&& \int d^8y\,
\Bigg(\frac{\delta^2\Gamma}{\delta\phi_x^*\delta\phi_y}\Bigg)^{-1}
\frac{D_y^2}{8\partial^2}
\frac{\delta^2\Gamma}{\delta\phi_y\delta\phi_z^*} + \int d^8y\,
\Bigg(\frac{\delta^2\Gamma}{\delta\phi_x^*\delta\tilde\phi_y^*}\Bigg)^{-1}
\frac{\bar D_y^2}{8\partial^2}
\frac{\delta^2\Gamma}{\delta\tilde\phi_y^*\delta\phi_z^*} =
-\frac{D_x^2}{2}\delta^8_{xz};\nonumber\\
&& \int d^8y\,
\Bigg(\frac{\delta^2\Gamma}{\delta\phi_x^*\delta\phi_y}\Bigg)^{-1}
\frac{D_y^2}{8\partial^2}
\frac{\delta^2\Gamma}{\delta\phi_y\delta\tilde\phi_z} + \int
d^8y\,\Bigg(\frac{\delta^2\Gamma}{\delta\phi_x^*\delta\tilde\phi_y^*}\Bigg)^{-1}
\frac{\bar D_y^2}{8\partial^2}
\frac{\delta^2\Gamma}{\delta\tilde\phi_y^*\delta\tilde\phi_z} = 0,
\end{eqnarray}

\noindent are

\begin{equation}\label{Inverse_Functions}
\Bigg(\frac{\delta^2\Gamma}{\delta\phi_x^*\delta\phi_y}\Bigg)^{-1}
= -  \frac{G D_x^2 \bar D_x^2}{4(\partial^2 G^2 + m^2 J^2)}
\delta^8_{xy};\qquad
\Bigg(\frac{\delta^2\Gamma}{\delta\phi_x\delta\tilde\phi_y}\Bigg)^{-1}
=  - \frac{m J \bar D_x^2}{\partial^2 G^2 + m^2 J^2}
\delta^8_{xy}.
\end{equation}

Nevertheless, there are also the following expressions

\begin{equation}
\frac{\delta}{\delta V_y} \frac{\delta}{\delta
\phi_1}\exp\Bigg(\frac{2}{i}\frac{\delta}{\delta
J_x}+2V_x\Bigg)\phi_x;\qquad\quad \frac{\delta}{\delta V_y}
\frac{\delta}{\delta \tilde
\phi_1^*}\exp\Bigg(\frac{2}{i}\frac{\delta}{\delta
J_x}+2V_x\Bigg)\phi_x.
\end{equation}

\noindent in Eq. (\ref{Main_Identity}). In order to calculate them
at zero momentum of the gauge field $V$, we need supersymmetric
Ward identities.

It is necessary to note, that from Schwinger-Dyson equations
(\ref{SD_For_Matter1}) and (\ref{SD_For_Matter2}) similarly to
identities (\ref{Exponents}) it is easy to find

\begin{eqnarray}\label{Vertex_Exponent1}
&& - \frac{D_x^2}{2}\frac{\delta^2\Gamma}{\delta V_y \delta\phi_1
\delta\phi_{0x}^*} = - \frac{D_x^2}{8}\frac{\delta}{\delta
V_y}\frac{\delta}{\delta \phi_1}
\exp\Bigg(\frac{2}{i}\frac{\delta}{\delta J_x}+2V_x\Bigg) \phi_x =
\frac{\delta^3\Gamma}{\delta V_y \delta\phi_1 \delta\phi_x^*};\quad\\
\label{Vertex_Exponent2} && -
\frac{D_x^2}{2}\frac{\delta^2\Gamma}{\delta V_y
\delta\tilde\phi_1^* \delta\phi_{0x}^*} = - \frac{D_x^2}{8}
\frac{\delta}{\delta V_y} \frac{\delta}{\delta \tilde \phi_1^*}
\exp\Bigg(\frac{2}{i}\frac{\delta}{\delta J_x}+2V_x\Bigg) \phi_x =
\frac{\delta^3\Gamma}{\delta V_y \delta\tilde \phi_1^* \delta
\phi_x^*}.\quad
\end{eqnarray}

%%%%%%%%%%%%%%%%%%%%%%%%%%%%%%%%%%%%%%%%%%%%%%%%%%%%%%%%%%%%%%%%%%%

\section{Ward identities for $N=1$ supersymmetric electrodynamics}
\label{Section_Ward}
\hspace{\parindent}

The Ward identities for the supersymmetric electrodynamics are
obtained by the standard way \cite{Slavnov_Book}: substitution
(\ref{Gauge_Transformations}) is performed in generating
functional (\ref{Modified_Z}), the result is differentiated with
respect to $A$ in the limit $A\to 0$, and then we pass from the
generating functional $Z$ to the functional $\Gamma$. As a result
of this operation we obtain the identity

\begin{eqnarray}\label{Ward_Identity0}
&& \Big(\bar D_x^2 + D_x^2\Big)\Bigg(\frac{\delta\Gamma}{\delta
V_x}+\frac{1}{2e_0^2}\partial^2
\Big(1+\frac{\partial^{2n}}{\Lambda^{2n}}\Big)V_x\Bigg) - \bar
D_x^2 \Bigg( 2 \phi_0\frac{\delta\Gamma}{\delta\phi_0} - 2
\tilde\phi_0\frac{\delta\Gamma}{\delta\tilde\phi_0} +
\frac{1}{4}\phi_x
\frac{D_x^2}{\partial^2}\frac{\delta\Gamma}{\delta\phi_x}
-\nonumber\\
&& - \frac{1}{4}\tilde\phi_x
\frac{D_x^2}{\partial^2}\frac{\delta\Gamma}{\delta\tilde\phi_x}\Bigg)
- D_x^2\Bigg(2 \phi_0^* \frac{\delta\Gamma}{\delta\phi_0^*} - 2
\tilde\phi_0^* \frac{\delta\Gamma}{\delta\tilde\phi_0^*} +
\frac{1}{4}\phi_x^* \frac{\bar
D_x^2}{\partial^2}\frac{\delta\Gamma}{\delta\phi_x^*} -
\frac{1}{4}\tilde\phi_x^* \frac{\bar
D_x^2}{\partial^2}\frac{\delta\Gamma}{\delta\tilde\phi_x^*}\Bigg)
= 0.\qquad
\end{eqnarray}

\noindent (in which the fields are not yet set equal to 0).
Differentiating this equality with respect to $\phi_y$ and
$\phi_{0z}^*$ and then setting all fields equal to 0, we find the
Ward identity for the three-point Green function

\begin{eqnarray}\label{Ward_Identity1}
(\bar D_x^2 + D_x^2) \frac{\delta^3\Gamma}{\delta V_x \delta\phi_y
\delta\phi_{0z}^*} = 2 \bar D_x^2 \delta^8_{xy}
\frac{\delta^2\Gamma}{\delta\phi_x \delta\phi_{0z}^*} + 2
D_x^2\Bigg( \delta^8_{xz} \frac{\delta^2\Gamma}{\delta\phi_y
\delta\phi_{0x}^*}\Bigg)
\end{eqnarray}

\noindent where we take into account that a variational derivative
with respect to a chiral superfield is a chiral superfield again.

Solution of Ward identity (\ref{Ward_Identity1}) in the limit
$p\to 0$ is a function

\begin{eqnarray}\label{Vertex1}
&& \frac{\delta^3\Gamma}{\delta
V_x\delta\phi_y\delta\phi^*_{0z}}\Bigg|_{p=0} = \frac{1}{4}
\frac{\delta}{\delta V_x}\frac{\delta}{\delta \phi_y}
\exp\Bigg(\frac{2}{i}\frac{\delta}{\delta J_z}+2V_z\Bigg)
\phi_z\Bigg|_{p=0} =\nonumber\\
&& = -2 \partial^2\Pi_{1/2}{}_x\Big(\bar D_x^2\delta^8_{xy}
\delta^8_{xz}\Big) F(q^2) + \frac{1}{8} D^a C_{ab} \bar
D_x^2\Big(\bar
D_x^2\delta^8_{xy} D_x^b \delta^8_{xz} \Big) f(q^2)
+\vphantom{\frac{1}{2}}\nonumber\\
&&\qquad\qquad\qquad\qquad -\frac{1}{16} q^\mu G'(q^2) \bar
D\gamma^\mu\gamma_5 D_x \Big(\bar D_x^2\delta^8_{xy}
\delta^8_{xz}\Big) -\frac{1}{4} \bar D_x^2\delta^8_{xy}
\delta^8_{xz}\, G(q^2),\qquad
\end{eqnarray}

\noindent where

\begin{equation}
\Pi_{1/2} = - \frac{1}{16 \partial^2} D^a \bar D^2 C_{ab} D^b = -
\frac{1}{16 \partial^2} D^a D^2 C_{ab} D^b
\end{equation}

\noindent is the supersymmetric transversal projector. $F(q^2)$
and $f(q^2)$ are some functions of the matter field momentum $q$,
which can not be determined from the Ward identity. Using Eq.
(\ref{Vertex_Exponent1}) we also obtain

\begin{eqnarray}\label{Vertex2}
&& \frac{\delta^3\Gamma}{\delta
V_x\delta\phi_y\delta\phi^*_{z}}\Bigg|_{p=0} =
\partial^2\Pi_{1/2}{}_x\Big(\bar D_x^2\delta^8_{xy} D_x^2
\delta^8_{xz}\Big) F(q^2) +\nonumber\\
&& \qquad\qquad\qquad +\frac{1}{32} q^\mu G'(q^2) \bar
D\gamma^\mu\gamma_5 D_x \Big(\bar D_x^2\delta^8_{xy} D_x^2
\delta^8_{xz}\Big) + \frac{1}{8} \bar D_x^2\delta^8_{xy} D_x^2
\delta^8_{xz}\, G(q^2),\qquad
\end{eqnarray}

Similarly, differentiating Eq. (\ref{Ward_Identity0}) with respect
to $\phi_y$ and $\tilde\phi_{0z}$, we obtain Ward identity

\begin{eqnarray}
&& (D_x^2 + \bar D_x^2) \frac{\delta^3\Gamma}{\delta V_x
\delta\phi_y \delta\tilde\phi_{0z}} = - \bar D_x^2 \Bigg(\bar
D_x^2 \delta^8_{xy} \frac{D_x^2}{8 \partial^2}
\frac{\delta^2\Gamma}{\delta\phi_x\delta\tilde\phi_{0z}} + 2
\delta^8_{xz}
\frac{\delta^2\Gamma}{\delta\tilde\phi_{0x}\delta\phi_y}\Bigg)
\end{eqnarray}

\noindent Taking into account that according to Eqs.
(\ref{Exponents}) and (\ref{Explicit_Green_Functions})

\begin{equation}
\frac{\delta^2\Gamma}{\delta\tilde\phi_{0x}\delta\phi_y} =
\frac{m}{32\partial^2} \Big(J(\partial^2)-1\Big) \bar D_y^2 D_y^2
\delta^8_{xy},
\end{equation}

\noindent it is possible to rewrite the considered Ward identity
in the following form:

\begin{eqnarray}
&& (D_x^2 + \bar D_x^2) \frac{\delta^3\Gamma}{\delta V_x
\delta\phi_y
\delta\tilde\phi_{0z}} =\nonumber\\
&&\qquad = \frac{m}{16} \Bigg[\bar D_x^2 \delta^8_{xy} \frac{\bar
D_x^2 D_x^2}{\partial^2} \Big(J(\partial^2)-1\Big) \delta^8_{xz} -
\bar D_x^2 \Bigg(\delta^8_{xz} \frac{D_x^2 \bar D_x^2}{\partial^2}
\Big(J(\partial^2)-1\Big) \delta^8_{xy}\Bigg)\Bigg].\qquad
\end{eqnarray}

\noindent Its solution at $p=0$ can be written as

\begin{eqnarray}\label{Vertex3}
&& \frac{\delta^3\Gamma}{\delta V_x \delta\phi_y
\delta\tilde\phi_{0z}}\Bigg|_{p=0} = \frac{1}{4}
\frac{\delta}{\delta V_x} \frac{\delta}{\delta\phi_y}
\exp\Bigg(\frac{2}{i}\frac{\delta}{\delta J_z}+2V_z\Bigg)
\tilde\phi_z^*\Bigg|_{p=0} =\nonumber\\
&& = -\frac{m}{2} \partial^2 \Pi_{1/2}{}_x \Bigg( \bar
D_x^2\delta^8_{xy} D_x^2 \delta^8_{xz} - D_x^2 \bar D_x^2
\delta^8_{xy} \delta^8_{xz} \Bigg) H(q^2)
+ \frac{m}{32} D^a C_{ab} D_x^2 \times\nonumber\\
&& \times  \Big(D_x^2 \bar D_x^2\delta^8_{xy} D_x^b
\delta^8_{xz}\Big) h(q^2) +\frac{m}{16} J'(q^2) \Bigg( \bar
D_x^2\delta^8_{xy} D_x^2 \delta^8_{xz} - D_x^2 \bar D_x^2
\delta^8_{xy} \delta^8_{xz} \Bigg)
+\nonumber\\
&& +\frac{m}{16}\Bigg(\frac{J'(q^2)}{q^2} - \frac{J(q^2)-1}{q^4}
\Bigg) \Bigg(D_x^2 \bar D_x^2\delta^8_{xy} \frac{\bar D_x^2
D_x^2}{16} \delta^8_{xz} + D_x^2 \bar D_x^2 \delta^8_{xy} q^2
\delta^8_{xz} \Bigg),
\end{eqnarray}

\noindent where $H(q^2)$ and $h(q^2)$ are two more functions,
which can not be found from the Ward identity. Here we take into
account, that the result should be chiral in $y$ and antisymmetric
with respect to the replacement $y\leftrightarrow z$ after
applying the operator $D_z^2$. Really, in this case from Eq.
(\ref{Vertex_Exponent2}) we also obtain

\begin{equation}
(D_x^2 + \bar D_x^2) \frac{\delta^3\Gamma}{\delta V_x \delta\phi_y
\delta\tilde\phi_{z}} = \frac{m}{2} \Bigg[\bar D_x^2 \delta^8_{xy}
\bar D_x^2 J(\partial^2) \delta^8_{xz} - \bar D_x^2 \delta^8_{xz}
\bar D_x^2 J(\partial^2) \delta^8_{xy}\Bigg].\qquad
\end{equation}

\noindent Solution of this equation in the limit $p\to 0$ is

\begin{eqnarray}\label{Vertex4}
&& \frac{\delta^3\Gamma}{\delta V_x \delta\phi_y
\delta\tilde\phi_{z}}\Bigg|_{p=0} = - \frac{\bar D_z^2}{2}
\frac{\delta^3\Gamma}{\delta V_x \delta\phi_y
\delta\tilde\phi_{0z}}\Bigg|_{p=0} = \nonumber\\
&& = \frac{m}{4} \partial^2 \Pi_{1/2}{}_x \Bigg( \bar
D_x^2\delta^8_{xy} D_x^2 \bar D_x^2 \delta^8_{xz} - D_x^2 \bar
D_x^2 \delta^8_{xy} \bar D_x^2\delta^8_{xz} \Bigg) H(q^2) -\nonumber\\
&&\qquad\qquad\qquad\qquad -\frac{m}{32} J'(q^2) \Bigg( \bar
D_x^2\delta^8_{xy} D_x^2 \bar D_x^2 \delta^8_{xz} - D_x^2 \bar
D_x^2 \delta^8_{xy} \bar D_x^2\delta^8_{xz} \Bigg).\qquad
\end{eqnarray}

%%%%%%%%%%%%%%%%%%%%%%%%%%%%%%%%%%%%%%%%%%%%%%%%%%%%%%%%%%%%%%%%%%%

\section{Two-point Green function of the gauge field}
\label{Section_Two_Point_Function}
\hspace{\parindent}

Let us now substitute the expressions, obtained above, into
two-point Green function of the gauge field (\ref{Main_Identity}).
Let us remind once again, that at first we make the calculation
ignoring diagrams, containing insertions of counterterms on lines
of the matter superfields. Then due to the supersymmetric gauge
invariance quantum corrections to the effective action,
corresponding to the two-point Green function of the gauge field,
can be written as

\begin{equation}\label{D0_Definition}
\Gamma^{(2)}_V = - \frac{1}{16\pi} \int
\frac{d^4p}{(2\pi)^4}\,V(-p)\,\partial^2\Pi_{1/2} V(p)\,
d_0^{-1}(\alpha_0,\Lambda/p).
\end{equation}

\noindent where $d_0$ is a function, which can be determined from
Eq. (\ref{Main_Identity}). Really, it is easy to see, that

\begin{eqnarray}\label{Two_Point_Function}
&& \Pi_{1/2}\, \int d^4x\,d^4y\,\frac{\delta^2 \Gamma}{\delta
V_x\,\delta V_y} \Bigg|_{V,\phi=0} \exp\Big(i p_\mu x^\mu + i
q_\mu y^\mu \Big)
=\nonumber\\
&&\qquad\qquad\qquad\qquad = \frac{1}{8\pi} (2\pi)^4
\delta^4\Big(p + q\Big)\, p^2 \Pi_{1/2}
\delta^4(\theta_x-\theta_y)\,d_0^{-1}(\alpha_0,\Lambda/p).\qquad
\end{eqnarray}

In order to find the function $d_0$ first we differentiate
expression (\ref{Main_Identity}) with respect to $\ln\Lambda$, and
then set $p=0$ in it. (We must take into account the dependence
$e_0 = e_0(\Lambda)$, which originates from diagrams with
insertions of the counterterms on lines of the gauge superfield
$V$.) Note, that due to the using of the higher derivative
regularization we can directly differentiate the integrand. The
differentiation with respect to $\ln\Lambda$ is needed in order to
obtain a finite expression at $p=0$. (Existence of the finite
limit at $p\to 0$ will be proven later.)

\begin{figure}[h]
\hspace*{4cm}
\includegraphics[scale=0.88]{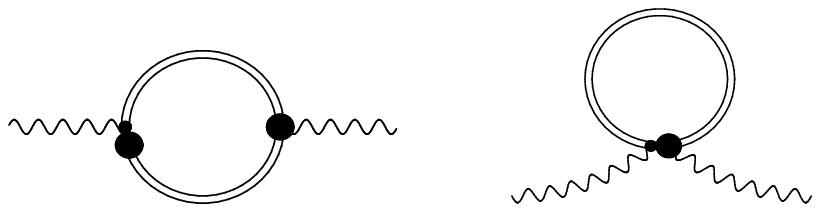}
\caption{Feynman diagrams, defining the $\beta$-function of $N=1$
supersymmetric electrodynamics.} \label{Figure_Effective_Diagrams}
\end{figure}

After differentiating Eq. (\ref{Main_Identity}) with respect to
$\ln\Lambda$ the contribution of the two-point Green function of
the gauge field to the effective action (without gauge fixing
terms) can be written as

\begin{equation}\label{Delta_Gamma}
\frac{d}{d\ln\Lambda} \Gamma^{(2)}_V = \frac{1}{2}
\frac{d}{d\ln\Lambda}\int d^8x\,d^8y\,V_x
V_y\Bigg(\Big(T^{(1)}_{xy} + T^{(2)}_{xy}\Big)-\sum\limits_i c_i
\Big(T^{(1)PV}_{xy}(M_i) + T^{(2)PV}_{xy}(M_i) \Big)\Bigg),
\end{equation}

\noindent where $T^{(1)}_{xy}$ denotes the sum of terms, which
correspond to the first diagram in Fig.
\ref{Figure_Effective_Diagrams}, and $T^{(2)}_{xy}$ is a sum of
diagrams, corresponding to the second diagram in this figure.
$T^{(1)PV}_{xy}(M_i)$ and $T^{(2)PV}_{xy}(M_i)$ in Eq.
(\ref{Delta_Gamma}) are the similar contributions of diagrams with
the Pauli-Villars fields.

After substitution vertex functions from Eqs. (\ref{Vertex1}),
(\ref{Vertex2}), (\ref{Vertex3}) and (\ref{Vertex4}) and
propagators from Eqs. (\ref{Explicit_Green_Functions}), Weak
rotation and some simple transformations, using the algebra of the
covariant derivatives, we find, that in the momentum
representation the first diagram is

\begin{eqnarray}\label{First_Diagram}
&& \frac{1}{2} \frac{d}{d\ln\Lambda}\int d^8x\,d^8y\,V_x
V_y\Bigg(T^{(1)}_{xy}
-\sum\limits_i c_i T^{(1)PV}_{xy}(M_i) \Bigg)=\nonumber\\
&& = \frac{d}{d\ln\Lambda} \int \frac{d^4p}{(2\pi)^4}
\frac{d^4q}{(2\pi)^4}\Bigg\{ V\partial^2\Pi_{1/2}V \Bigg[\frac{8 G
F}{q^2 G^2+m^2 J^2} - \frac{m^2 J J'}{2 q^2(q^2 G^2
+ m^2 J^2)}  +\nonumber\\
&& + \frac{1}{2}\frac{d}{dq^2}\Bigg(\ln\Big(q^2 G^2 + m^2 J^2\Big)
+ \frac{m^2 J}{q^2 G^2 + m^2 J^2}\Bigg)\Bigg] + V^2 \frac{G^2}{q^2
G^2 + m^2 J^2}\Bigg\},
\end{eqnarray}

\noindent while the second one is

\begin{eqnarray}\label{Second_Diagram}
&& \frac{1}{2} \frac{d}{d\ln\Lambda}\int d^8x\,d^8y\,V_x
V_y\Bigg(T^{(2)}_{xy}
-\sum\limits_i c_i T^{(2)PV}_{xy}(M_i) \Bigg)=\nonumber\\
&& = \frac{d}{d\ln\Lambda} \int \frac{d^4p}{(2\pi)^4}
\frac{d^4q}{(2\pi)^4}\Bigg\{V\partial^2\Pi_{1/2}V \Bigg[-\frac{8 G
F}{q^2 G^2+m^2 J^2} - \frac{m^2 JJ'}{2 q^2(q^2G^2+m^2J^2)}
+\nonumber\\
&& + \frac{m^2 J (J-1)}{2 q^4 (q^2 G^2+m^2 J^2)} - \frac{8 G f + 8
m^2 J h}{q^2 G^2 + m^2 J^2} \Bigg] - V^2 \frac{G^2}{q^2 G^2 + m^2
J^2}\Bigg\}.
\end{eqnarray}

\noindent Adding the results, we find

\begin{eqnarray}\label{I_Definition}
&& \frac{d}{d\ln\Lambda} \Gamma^{(2)}_V = \int
\frac{d^4p}{(2\pi)^4} V(-p)\,\partial^2 \Pi_{1/2} V(p)\, I,\qquad
\end{eqnarray}

\noindent where $I$ denotes the following Euclidean integral:

\begin{eqnarray}\label{Result}
&& I = \frac{d}{d\ln\Lambda}\int\frac{d^4q}{(2\pi)^4}\Bigg\{
\frac{1}{2q^2}\frac{d}{dq^2} \Bigg(\ln\Big(q^2 G^2 + m^2
J^2\Big) + \frac{m^2 J}{q^2 G^2 + m^2 J^2}\Bigg) -\nonumber\\
&& \qquad\qquad\qquad\qquad\qquad\qquad -\frac{m^2 J\,\Big(2 q^2
J' - (J-1)\Big)}{2 q^4\Big(q^2 G^2 + m^2 J^2\Big)}  - \frac{8 G f
+ 8 m^2 J h}{q^2 G^2 + m^2 J^2}
-\qquad \nonumber\\
&& - \sum\limits_i c_i \frac{1}{2q^2}\frac{d}{dq^2}
\Bigg(\ln\Big(q^2 G_{PV}^2 + M_i^2 J_{PV}^2\Big) + \frac{M_i^2
J_{PV}}{q^2 G_{PV}^2 + M_i^2 J_{PV}^2}\Bigg) +\nonumber\\
&& \qquad\qquad\qquad\qquad +\sum\limits_i c_i \Bigg(\frac{M_i^2
J\,\Big(2 q^2 J' - (J-1)\Big)}{2 q^4\Big(q^2 G^2 + M_i^2 J^2\Big)}
+ \frac{8 G f + 8 M_i^2 J h}{q^2 G^2 + M_i^2
J^2}\Bigg)\Bigg\}.\qquad
\end{eqnarray}

Thus we see, that all noninvariant terms, proportional to $V^2$,
and also terms, containing the undefined functions $F$ and $H$,
are completely cancelled. Nevertheless the final result contains
arbitrary functions $f$ and $h$, which can not be found from the
Ward identities. (In the two-loop approximation these functions
are equal to 0. They are nontrivial only at the three loops.)

From Eqs. (\ref{First_Diagram}) and (\ref{Second_Diagram}) we see,
that for the first diagram in Fig. \ref{Figure_Effective_Diagrams}
the result is obtained exactly to all orders of the perturbation
theory. In the three-loop approximation it was checked explicitly.
The expression for the second diagram contains the arbitrary
functions $f$ and $h$. However, these functions can be possibly
determined from other arguments, because the explicit three-loop
calculations, made in \cite{ThreeLoop}, enable us to propose the
following structure of the result:

\begin{eqnarray}\label{Proposal}
&& I = \frac{d}{d\ln\Lambda}\int\frac{d^4q}{(2\pi)^4}
\frac{1}{2q^2}\frac{d}{dq^2}
\Bigg\{\Bigg(\ln\Big(q^2 G^2 + m^2 J^2\Big) + X(q^2,m^2)\Bigg) -\nonumber\\
&& \qquad\qquad\qquad\qquad - \sum\limits_i c_i \Bigg(\ln\Big(q^2
G_{PV}^2 + M_i^2 J_{PV}^2\Big) + X(q^2,M_i^2)\Bigg)\Bigg\},
\end{eqnarray}

\noindent where $X(q^2,m^2)$ is a function, which is finite and
mass independent constant at $q=0$, and is equal to 0 in the limit
$q\to\infty$. If this proposal is true, the integral over the
momentum $q$ is reduced to the integral from a total derivative
and can be taken analytically. It will be made in the next
section.

\section{Renormgroup functions in N=1 supersymmetric electrodynamics.}
\label{Section_Anomaly_Puzzle} \hspace{\parindent}

Let us believe, that proposal (\ref{Proposal}) is true and find
its consequences. For this purpose let us pass to the
four-dimensional spherical coordinates and take the integral:

\begin{eqnarray}\label{Proposal_Integral}
&& I = \frac{1}{32\pi^2}\frac{d}{d\ln\Lambda}
\Bigg\{\Bigg(\ln\Big(q^2 G^2 + m^2 J^2\Big) + X(q^2,m^2)\Bigg) -\nonumber\\
&& \qquad\qquad\qquad\qquad - \sum\limits_i c_i \Bigg(\ln\Big(q^2
G_{PV}^2 + M_i^2 J_{PV}^2\Big) +
X(q^2,M_i^2)\Bigg)\Bigg\}\Bigg|_0^\infty.
\end{eqnarray}

Due to the higher derivative regularization the functions $G(q)$,
$J(q)$, $G_{PV}(q)$ and $J_{PV}(q)$ are equal to 1 in the limit
$q\to \infty$, because all quantum corrections to the classical
values disappear in this limit. Really, with the higher
derivatives all integrals are convergent at finite $\Lambda$ and
the momentum $q$ is present in denominators of integrals, defining
quantum corrections. For example,

\begin{equation}
\lim\limits_{q\to\infty} \int
\frac{d^4k}{(2\pi)^4}\,\frac{1}{\Big(1+k^{2n}/\Lambda^{2n}\Big)\,k^2(k+q)^2}
= 0.
\end{equation}

\noindent Therefore, taking into account that $\sum\limits_i c_i =
1$, the substitution at $q\to\infty$ in Eq.
(\ref{Proposal_Integral}) is equal to 0.

Below we will consider massive and massless cases separately.

{\bf 1. Massless case:}

Taking into account, that the function $J_{PV}$ is a finite, mass
and $\Lambda$ independent constant in the limit $q\to 0$, because
it is defined by convergent (even in the limit
$\Lambda\to\infty$), dimensionless integrals, which do not contain
infrared divergences, we obtain

\begin{equation}\label{Limit}
I = - \frac{d}{d\ln\Lambda} \frac{1}{32\pi^2}\Bigg\{\ln G^2 -
\sum\limits_i c_i \ln\Big(M_i^2 J_{PV}^2\Big)\Bigg\}\Bigg|_{q=0} =
\frac{1}{16\pi^2}\Bigg(1-\frac{d\ln G}{d\ln
\Lambda}\Bigg)\Bigg|_{q=0}.
\end{equation}

\noindent Now let us prove, that such limit exists, because the
result in principle can diverge at $q\to 0$. First we will
suppose, that all terms, which gives 0 in the limit $\Lambda\to
\infty$, are removed from the function $G(\alpha_0,q/\Lambda)$.
(For example, the terms, proportional to $(q/\Lambda)^k$, where
$k$ is a positive constant.) It is evident, that removing of such
terms does not change renormgroup functions. Because in this case
$ZG$ is finite due to the definition of the renormalization
constant $Z$ and does not depend on $\Lambda$, then

\begin{eqnarray}
&& \Lambda \frac{d}{d\Lambda} \ln
G\Big(\alpha_0(\alpha,\Lambda/\mu),\Lambda/p\Big)\Bigg|_{p=0}
=\nonumber\\
&&\qquad\qquad\quad = -\Lambda \frac{d}{d\Lambda} \ln
Z\Big(\alpha_0(\alpha,\Lambda/\mu),\Lambda/p\Big)\Bigg|_{p=0} =
-\gamma\Big(\alpha_0(\alpha,\Lambda/\mu)\Big).\qquad
\end{eqnarray}

\noindent The anomalous dimension $\gamma(\alpha_0)$ is finite at
$\Lambda<\infty$, so that limit (\ref{Limit}) exists in all orders
of the perturbation theory.

From Eqs. (\ref{D0_Definition}), (\ref{I_Definition}) and
(\ref{Limit}) we obtain

\begin{equation}
\frac{d}{d\ln\Lambda}
d_0^{-1}\Big(\alpha_0,\Lambda/p\Big)\Bigg|_{p=0} = \frac{1}{\pi}
\frac{d}{d\ln\Lambda}\Bigg(\ln\frac{\Lambda}{p} -
G\Big(\alpha_0,\Lambda/p\Big)\Bigg)\Bigg|_{p=0}.
\end{equation}

\noindent This equation defines the function $d_0$ up to an
arbitrary integration constant, because this function is a
polynomial in $\ln\Lambda/p$. Therefore, due to Eq.
(\ref{D0_Definition}) the contribution of the two-point Green
function of the gauge superfield to the effective action without
diagrams, containing insertions of counterterms on lines of the
matter superfields, is

\begin{equation}\label{Z=1_Result}
\int \frac{d^4p}{(2\pi)^4}\,V(-p)\,\partial^2 \Pi_{1/2}
V(p)\,\Bigg[- \frac{1}{4e_0^2} - \frac{1}{16\pi^2}
\Big(\ln\frac{\Lambda}{p} - \ln G(\alpha_0,\Lambda/p) \Big) +
\mbox{const}\Bigg].
\end{equation}

However it is also necessary to take into account diagrams with
insertions of counterterms. Because with the Pauli-Villars fields
there are no divergences in the theory at a finite $\Lambda$, it
is possible to perform the rescaling $\phi\to Z^{-1/2}\phi$,
$\tilde\phi\to Z^{-1/2}\tilde\phi$. However this rescaling should
be complemented by the simultaneous rescaling of Pauli-Villars
fields $\Phi\to Z^{-1/2}\Phi$, $\tilde\Phi\to Z^{-1/2}\tilde\Phi$
in order to avoid appearance of the divergences in diagrams with
insertions of counterterms. (Or, equivalently, in order to avoid
anomalous contribution of the functional integral measure.) It is
easy to see, that such rescaling is actually reduced to the
substitution $m\to m/Z$, $M_i\to M_i/Z$. Making this substitution
in Eq. (\ref{Limit}) and taking into account that at $Z=1$ the
result is given by Eq. (\ref{Z=1_Result}), at $Z\ne 1$ we obtain

\begin{equation}
\Gamma_V^{(2)} = \int
\frac{d^4p}{(2\pi)^4}\,V(-p)\,\partial^2\Pi_{1/2} V(p)\,
\,\Bigg[-\frac{1}{4e_0^2} - \frac{1}{16\pi^2} \ln\frac{\Lambda}{p}
+ \frac{1}{16\pi^2}\ln (Z G) + \mbox{const}\Bigg].
\end{equation}

Because $ZG$ is finite according to the definition of the
renormalization constant $Z$, then, in order to make this
expression finite, it is necessary to compensate only the one-loop
divergence. For this purpose the bare charge is presented as

\begin{equation}\label{Bare_Coupling_Constant}
\frac{1}{e^2} Z_3\Big(e,\Lambda/\mu\Big) = \frac{1}{e_0^2} =
\frac{1}{e^2} - \frac{1}{4\pi^2}\ln\frac{\Lambda}{\mu} +
\mbox{const}.
\end{equation}

\noindent Therefore, the final expression for the effective action
(without gauge fixing terms) in the massless case can be written
as

\begin{equation}\label{Effective_Action}
\Gamma_V^{(2)} = \int \frac{d^4p}{(2\pi)^4}\,W_a(-p)\,C^{ab}
W_b(p) \,\Bigg[\frac{1}{4e^2} + \frac{1}{16\pi^2} \ln\frac{\mu}{p}
- \frac{1}{16\pi^2}\ln (Z G) + \mbox{const}\Bigg].
\end{equation}

Differentiating Eq. (\ref{Bare_Coupling_Constant}) with respect to
$\ln\mu$ at fixed value of the bare coupling constant $e_0$, we
find, that the $\beta$-function, defined by

\begin{equation}\label{Beta_Definition}
\beta = \frac{d}{d\ln\mu}\Bigg(\frac{e^2}{4\pi}\Bigg),
\end{equation}

\noindent is

\begin{equation}\label{One_Loop_Beta}
\beta(\alpha) = \frac{\alpha^2}{\pi}.
\end{equation}

\noindent Such result completely agrees with the structure of the
anomalies supermultiplet, which implies, that this
$\beta$-function is exhausted at the one-loop.

Moreover it is possible to define Gell-Mann-Low $\beta$-function
as follows: If

\begin{equation}\label{D_Definition}
\Gamma_V^{(2)} = \frac{1}{16\pi} \int
\frac{d^4p}{(2\pi)^4}\,W_a(-p)\,C^{ab} W_b(p)
\,d^{-1}(\alpha,\mu/p).
\end{equation}

\noindent (unlike Eq. (\ref{D0_Definition}) this expression
includes contributions of diagrams with insertions of
counterterms), then we define

\begin{equation}\label{GL_Function}
\tilde\beta\Big(d(\alpha,x)\Big) \equiv
-\frac{\partial}{\partial\ln x}\, d(\alpha,x)\Bigg|_{x=1}.
\end{equation}

\noindent According to Eq. (\ref{Effective_Action})

\begin{equation}
d^{-1}(\alpha,\mu/p) = \frac{1}{\alpha} + \frac{1}{\pi}
\ln\frac{\mu}{p} - \frac{1}{\pi}\ln (Z G) + \mbox{const}.
\end{equation}

\noindent Differentiating this expression with respect to
$\ln\mu$, we obtain

\begin{equation}
\tilde\beta(\alpha) =
\frac{\alpha^2}{\pi}\Big(1-\tilde\gamma(\alpha)\Big),\quad\mbox{where}\quad
\tilde\gamma \equiv \frac{\partial}{\partial\ln x}\,\ln(ZG),
\end{equation}

\noindent where we take into account, that the product $ZG$
depends on $\alpha$ and $x\equiv \mu/p$. We see, that, unlike
$\beta$, $\tilde\beta$ has corrections in orders of the
perturbation theory.

{\bf 2. Massive case:}

In the massive case we will take into account diagrams with
insertions of counterterms on lines of the matter superfields from
the beginning. Therefore

\begin{eqnarray}
&& I = - \frac{1}{32\pi^2} \frac{d}{d\ln\Lambda}
\Bigg\{\Bigg(\ln\Big(q^2 G^2 + m^2 J^2/Z^2\Big)
+ X(q^2,m/Z)\Bigg)-\nonumber\\
&& - \sum\limits_i c_i \Bigg(\ln\Big(q^2 G_{PV}^2 + M_i^2
J_{PV}^2/Z^2\Big) + X(q^2,M_i/Z)\Bigg) \Bigg\}\Bigg|_{q=0} =
\frac{1}{16\pi^2},\qquad
\end{eqnarray}

\noindent where e.f. $G=G(q,m/Z,M_i/Z)$, if $G(q,m,M_i)$ is a
result of calculation of diagrams without insertions of
counterterms on lines of the matter superfields e.t.c. Then from
Eq. (\ref{I_Definition}) we find

\begin{equation}
\Gamma_V^{(2)} = \int
\frac{d^4p}{(2\pi)^4}\,V(-p)\,\partial^2\Pi_{1/2} V(p)\,
\,\Bigg[-\frac{1}{4e_0^2} - \frac{1}{16\pi^2} \ln\frac{\Lambda}{p}
+ \mbox{finite terms}\Bigg].
\end{equation}

\noindent Because, as earlier, this expression diverges (at fixed
$e_0$) only in the one-loop approximation, in the massive case the
bare charge is also presented in form
(\ref{Bare_Coupling_Constant}) and $\beta$-function
(\ref{Beta_Definition}) is given by Eq. (\ref{One_Loop_Beta}). As
earlier this result completely agree with the structure of the
anomalies supermultiplet and with the Adler-Bardeen theorem.

Note, that in the massive case it is not necessary to discuss
Gell-Man-Low function, because defining this function the masses
are considered to be negligibly small and the original massive
theory is actually changed by the massless one.

%%%%%%%%%%%%%%%%%%%%%%%%%%%%%%%%%%%%%%%%%%%%%%%%%%%%%%%%%%%%%%%%%%%

\section{Conclusion}
\label{Section_Conclusion}
\hspace{\parindent}

In this paper some contributions to the two-point Green function
of the gauge field at $p\to 0$ were calculated using
Schwinger-Dyson equations and Ward identities for the $N=1$
supersymmetric electrodynamics. Unfortunately we did not manage to
find this function completely from Ward identities, but the most
involved diagrams had been calculated completely.

Nevertheless, using the results of explicit three-loop
calculations it is possible to suggest structure of terms, which
can not be found from the Ward identity. Then the $\beta$-function
and Gell-Mann-Low function can be found exactly to all orders of
the perturbation theory. The obtained results allow to give a
simple and clear solution of the anomaly puzzle in the $N=1$
supersymmetric electrodynamics, which, in particular, relates
different solutions of the anomaly puzzle known in the literature
\cite{ThreeLoop}. Let us remind its main points:

Because the axial anomaly and the anomaly of the energy-momentum
tensor trace are components of a single supermultiplet,
$\beta$-function (\ref{Beta_Definition}) should be exhausted at
the one-loop according to Adler-Bardeen theorem \footnote{With the
higher derivative regularization this theorem is valid, while with
the dimensional reduction it seems to fail \cite{Kazakov}.} and is
given by Eq. (\ref{One_Loop_Beta}). This is the result, that is
found with the higher derivative regularization both for the
massless and for the massive theory. Therefore, the obtained
results completely agree with the structure of the anomalies
supermultiplet.

Nevertheless, Gell-Mann-Low function, which is defined by the
transversal part of the two-point Green function of the gauge
field, has corrections in all orders and coincides with the exact
NSVZ $\beta$-function. Note, that it is sensible to define
Gell-Mann-Low function only for the massless theory, because in
order to construct it, it is necessary to neglect the dependence
of the effective action on the mass.

Two $\beta$-functions, defined by the different ways, are
different because generating functional (\ref{Modified_Z}) depends
on the normalization point $\mu$ at fixed bare coupling constant
$e_0$ \cite{HD_And_DRED,ThreeLoop}. The matter is that two
definitions, given above, are equivalent only if such dependence
is absent. Really, if the generating functional does not depend on
$\mu$, the function $d(\alpha,\Lambda/\mu)$, defined by Eq.
(\ref{D_Definition}), also will not depend on $\mu$.
Differentiating this function with respect to $\ln\mu$, we obtain

\begin{equation}
0 = \tilde\beta\Big(d(\alpha,x)\Big) - \beta(\alpha)
\frac{\partial}{\partial\alpha} d(\alpha,x).
\end{equation}

\noindent In particular, at $x=1$

\begin{equation}\label{GL_And_Beta}
\tilde\beta(\tilde\alpha) =
\beta(\alpha)\frac{d\tilde\alpha}{d\alpha},
\end{equation}

\noindent where $\tilde\alpha\equiv d(\alpha,1)$. It means, that
two different definitions of the $\beta$-function are equivalent.
However in the considered case generating functional
(\ref{Modified_Z}) depends on the normalization point $\mu$ at the
fixed $e_0$. Really, the dependence on $\mu$ in Eq.
(\ref{Modified_Z}) comes from the $\mu$-dependent renormalization
constant $Z$. One can try to remove this dependence by the
substitution $\phi\to Z^{-1/2}\phi$. However, according to
\cite{Arkani} this substitution has an anomalous Jacobian, which
contains $\ln Z$, depending on $\mu$. (This Jacobian was
explicitly written in \cite{HD_And_DRED,ThreeLoop}.) Therefore,
even after the substitution $\phi\to Z^{-1/2}\phi$ generating
functional (\ref{Modified_Z}) will contain $\ln Z$ and, therefore,
will depend on $\mu$. As a consequence Eq. (\ref{GL_And_Beta}),
which relates $\beta$-functions defined by the different ways now
is not valid, and the functions $\beta$ and $\tilde\beta$ are not
equivalent in the considered case.

In principle it is possible to define the generating functional,
which does not depend on the normalization point. Two different
ways to do it were proposed in \cite{ThreeLoop}. Then
$\beta$-function (\ref{Beta_Definition}) has corrections in all
orders of the perturbation theory, but the arguments, based on the
structure of the anomalies supermultiplet fail due to some
reasons. That is why the anomaly puzzle does not take place if the
theory is regularized by higher derivatives.

Finally we mention, that it would be interesting to perform the
similar investigation for the supersymmetric Yang-Mills theory.
However for this theory the calculations with the higher covariant
derivative regularization are very involved, because the presence
of the term with higher {\it covariant} derivatives
\cite{West_Paper} essentially complicates the form of the action.
In this case the calculations can be considerably simplified if
the covariant derivatives are substituted by the ordinary
derivatives. Then the gauge invariance will be certainly broken,
so that Ward identities can be also broken by some local
noninvariant terms. Nevertheless the gauge invariance can be
restored by the special choice of the subtraction scheme, so that
it is possible to use noninvariant regularizations
\cite{Slavnov1,Slavnov2} for the calculations. Such scheme was
proposed in \cite{SlavnovStepan1} for Abelian supersymmetric
theories and in \cite{SlavnovStepan2} for the nonabelian theories.
So, it is possible, that application of the above described
regularization to supersymmetric Yang-Mills theories enables us to
perform similar calculations also in the nonabelian case. At
present this work is in progress.

\bigskip
\bigskip

\noindent {\Large\bf Acknowledgments.}

\bigskip

\noindent This work was supported by Russian Foundation for Basic
Research (grant No 02-01-00126).

%%%%%%%%%%%%%%%%%%%%%%%%%%%%%%%%%%%%%%%%%%%%%%%%%%%%%%%%%%%%%%%%%%%%%%%%%

\end{document}